# Attention U-net approach in predicting Intensity Modulated Radiation Therapy dose distribution in brain glioma tumor

Mobina Naeemi[1], Mohamad Reza Esmaeili [2], Iraj Abedi[3]

1. Department of Physics, Isfahan University of Technology, Isfahan, Iran
2. Department of Electrical and Computer Engineering, Isfahan University of Technology, Isfahan, Iran
3. Department of Medical Physics, School of Medicine, Isfahan University of Medical Science, Isfahan, Iran

m.naeemi@ph.iut.ac.ir, mohamadreza_esmaeili@alumni.iut.ac.ir, i.abedi@med.mui.ac.ir

Corresponding E-mail: i.abedi@med.mui.ac.ir

**Abstract**

Today, intensity modulated radiation therapy (IMRT) is one of the methods used to treat brain tumors. In conventional treatment planning methods, after identifying planning target volume (PTV), organs at risk (OARs), and determining the limitations for them to receive radiation, dose distribution is performed based on optimization algorithms, which is usually a time-consuming method. In this article, artificial intelligence is used to acquire the knowledge used in the treatment planning of past patients, to plan for new patients in order to speed up the process of treatment planning and determination the appropriate dose distribution. In this paper, using deep learning algorithms, two different approaches are studied to predict dose distribution and compared with actual dose distributions. In the first method, only the images containing PTV and the distribution of the corresponding doses are used to train the convolutional neural network, but in the second one, in addition to PTV, the contours of four OARs are also used to introduce the network. The results show that the performance of both methods on test patients have high accuracy and in comparison with each other almost have the same results and high speed to design the dose distribution. Due to the fact that the "Only-PTV" method does not have the process of OARs identifying, applying it in designing the dose distribution will be much faster than using the "PTV-OARs" method in the whole of treatment planning.

# 1. Introduction

Brain cancers arise in a region with vital structures, which has become very challenging due to the proximity of organs at risk of cancer, the presence of intersecting forms, and the need for variable dose levels. Technology advancements in radiation therapy have improved the treatment of this sensitive area. First, is intensity-modulated radiation therapy (IMRT), which uses different radiation intensities to treat tumors in complex areas [1-3]. The purpose of this technique is to send a dose to the cancerous tumor with the least damage to sensitive organs [4]. IMRT, with a reverse programming strategy, describing the desired dose distribution and extracting the beam intensity profile, has been widely used to treat various extracranial tumors [5]. The second technique is volumetric modulated arc therapy (VMAT) which dynamically administers the dose in a spiral pattern around the patient, which is less time-consuming than IMRT [6, 7]. Both methods rely on imaging (CT, MRI) for planning, significantly improving tumor targeting and dose delivery. However, the increasing complexity of the delivery method has led to more time spent on dosimetry designs [8]. This means that a computerized treatment plan is created by considering a collection of targets and restrictions for covering the tumor and preserving healthy tissue, and the software uses them to generate a large number of beam segments to deliver the required dose [9, 10]. Although the inverse planning process relies on computers, it requires human resources. A high level of treatment planner intervention is usually necessary to produce a high-quality plan [11]. The standard process of Dose-volume histograms (DVHs) and Isodose distributions became essential tools for plan evaluation. Together with the progress in image processing, information from CT also enabled accurate dose calculation using the convolution-superposition method, allowing the inhomogeneous distribution of tissues to be more accurately handled [12].

In recent years, with the rise of deep learning and its multiple applications in computer vision through convolutional neural networks (CNN), the use of machine learning in radiation oncology has grown exponentially [13-15]. Rosenberger et al. proposed a U-net architecture for two-dimensional (2D) medical image segmentation, which allows the recovery of spatial details due to the encoder-decoder architecture with long-hop connections during the sampling process [16]. In the following years, 3D images were developed to increase the availability of volumetric medical imaging data. This development was very effective due to correct dose prediction in 2D slices and avoiding significant errors near the boundaries of the planning target volume [17, 18]. On the other hand, despite the success of U-Net, it has limitations, including the loss of spatial information and the problem of image accuracy with changes in the lesion or tumor [16]. Furthermore, U-Net architectures suffer from three primary forms. First, it enables frequent transmission of low-resolution information to improve learning performance, which leads to the blurring of extracted image features. Second, the network does not provide high-resolution edge information, which leads to uncertainty in network decisions. Third, the number of integration operations depends on the object's size [19-22]. Oktay et al. and Schempler et al. presented a modular of attention gates (AG) in the standard U-net architecture (Attention U-net) developed for medical image segmentation tasks, which results in a memory-efficient attention module for U-net [20, 23]. This led to improved results and demonstrated the advantages of AG for identifying and localizing specific structures [19].

The drop in mean squared error (MSE) tendency to converge quickly regardless of the target domain makes it a prime candidate for most serious learning regression problems. However, its inability to acquire domain-specific knowledge remains a substantial limitation for medical imaging tasks, especially when dealing with dose prediction, which is still heavily based on knowledge-based and physics-based approaches [24]. In [14, 25], a dose planning method based on adversarial learning was proposed, which was sensitive to many meta-parameters and made training challenging. Nguyen et al. proposed a dosimetric target based on the dose-volume histogram, a standard metric and a deterministic tool for cancer treatment. They adapted the DVH metric to a different target and included an adversarial target to obtain residual domain knowledge. They observed significant improvements in dose estimation for prostate cancer, mainly in the form of reduced trade-offs between planning target volume (PTV) coverage and planning target volume reductions (OARs) [13].

In this paper, we propose a prediction model based on the Attention U-Net algorithm for dose prediction in brain cancer tumors, increasing the accuracy by using the existing knowledge of the IMRT method. Also, in this study, dose prediction using it based on DVH and comparison with Only-PTV and PTV-OARs methods are tested. We compare this model with the standard U-Net model in [16].

## 2. Materials and methods

### 2.1 Patient data

99 patients with brain cancer tumors who referred for IMRT treatment between 2018 and 2020 were collected in this research. All collected patients passed the clinical stage T13N0M0. CT images of collected patients were taken in a style with a thermoplastic mask system. To better define the target and OARs, patients' MRI is indistinguishable from their CT scan, and were combined by the PROWESS Panter treatment planning system (version 5.5). Gross tumor volume (GTV) was defined on T1-weighted MRI after gadolinium. The clinical target volume (CTV) is calculated by adding a 1 cm 3D dilation to the GTV. On the other hand, we increased the CTV by 1 cm to create the PTV. The curved OARs included the brainstem, optic nerve, optic chiasm, lenses, eyeball, and intact brain. All patients were planned manually and the sixth field IMRT was used for their treatment. The photon energy was 6 megavolts and the gate angles were 60°, 100°, 165°, 195°, 260° and 300°, respectively. All treatment plans were performed with the collapsed cone convolution (CCC) algorithm to provide 60 Gy over 30 fractions. All vital structures including chiasm, optic nerves, eyeball, brainstem and lenses can be considered as sensitive organs. As a result, planning limitations have focused on the maximum dose. A summary of dosimetry constraints and priorities leading to planning optimization is reported in Table 1. For each plan, the PTV and OAR contours were determined by experienced radiation oncologists, and the dose distribution was optimized and confirmed by experienced medical physicists.

Table 1. Dosimetrist constraints

| Organ at risk | Objective(s) |
| --- | --- |
| Brainstem | $D_{max} \leq 54 Gy$ |
| Cochlea | $D_{mean} \leq 45 Gy$ |
| Cortex | $D_{max} \leq 28.6 Gy$ |
| Eyes | $D_{max} \leq 45 Gy$ |
| Lens | $D_{max} \leq 6 Gy$ |
| Optic nerves | $D_{max} \leq 54 Gy$ |

**2.2 Proposed model**

This paper applies a self-attention gating model that features CNN-based standard image analysis models for dense label predictions. On the other hand, we combined the attention approach proposed in [22] with a network-based gating proposal that allows attention coefficients to be more specific for local areas with the model [20] to improve the model's sensitivity to foreground pixels without the need for complex heuristics. This improves performance compared to global feature vector-based gating. The accuracy improvement over U-Net has been empirically observed to be consistent across different imaging datasets. In addition, the proposed hybrid model can be used for dense predictions because we do not perform adaptive integration. We present one of the first cases using an attention algorithm in brain cancer tumor dose prediction. This is the first work in brain cancer medical imaging. To train the model, the similarity between the predicted doses and the actual doses is evaluated by the dice reduction function. Then the Adaptive Moment Estimation (ADAM) algorithm is used to optimize the weights or parameters during the training process.

It is noteworthy that in the two methods, in addition to PTV, OARs, and dose distribution, the scalp has also been used to determine the location of these elements in different slices of the brain image. The first method trains the network using two-channel images (white and black) at the input. But in the second method to separate OARs and PTVs from each other, each element is marked with a specific color and is introduced to the network in three-channel images (RGB mode). On the other hand, the doses with different percentages were trained independently on the web, and the images of the input sections were used as targets. In this article, the images of 90 patients were used for the training and validation set, and the remaining nine patients were used to evaluate the network performance. . The images of the patients for the training and test set were done randomly. The IMRT images collected by the treatment planning system were obtained in DICOM format. Different lines of OAR and dose distribution in BMP format. We extracted by DICOM TO MATLAB. We used many images of further cuts to increase the clarity of the image in training; for this reason, this practice increases the memory during network training, which is all images with a size of 64 x 64 x 3 for three-channel images and 64 x 64 for two-channel images. The channel is done.

In addition, In this study, we used two proposed methods. The first method, all the images related to a patient, which contained only PTV, were selected for the training set and applied as input to the algorithm (Figure 1). Second method, in addition to PTV, we selected images with OAR as input images (Figure 2). We trained the dose related to these images as the target of the network. On the other hand, more than 5000 image cuts have been used for each training set, so that the predicted dose for different percentages of the dose can be displayed independently.

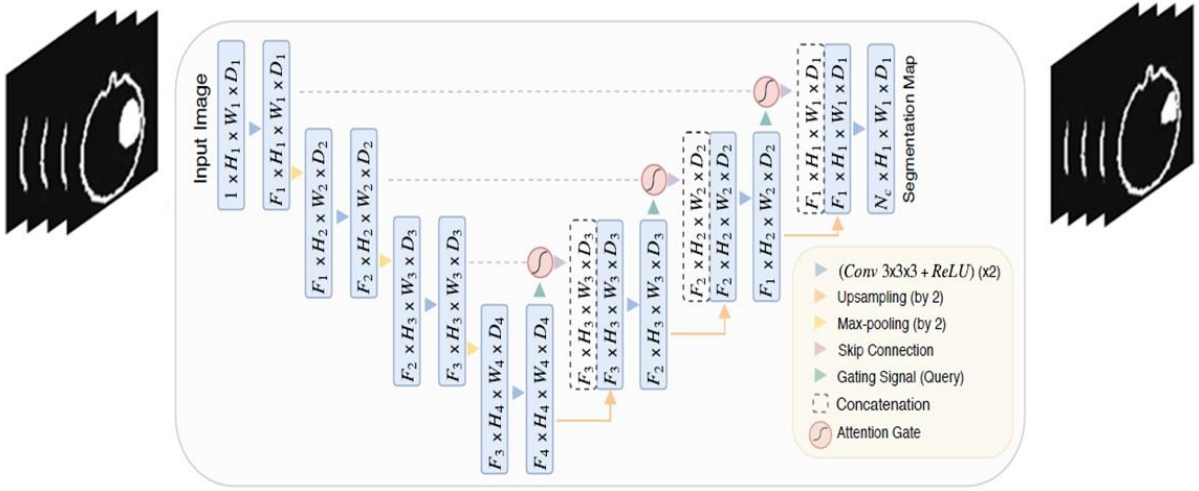

**Fig 1:** the first proposed method of network training structure

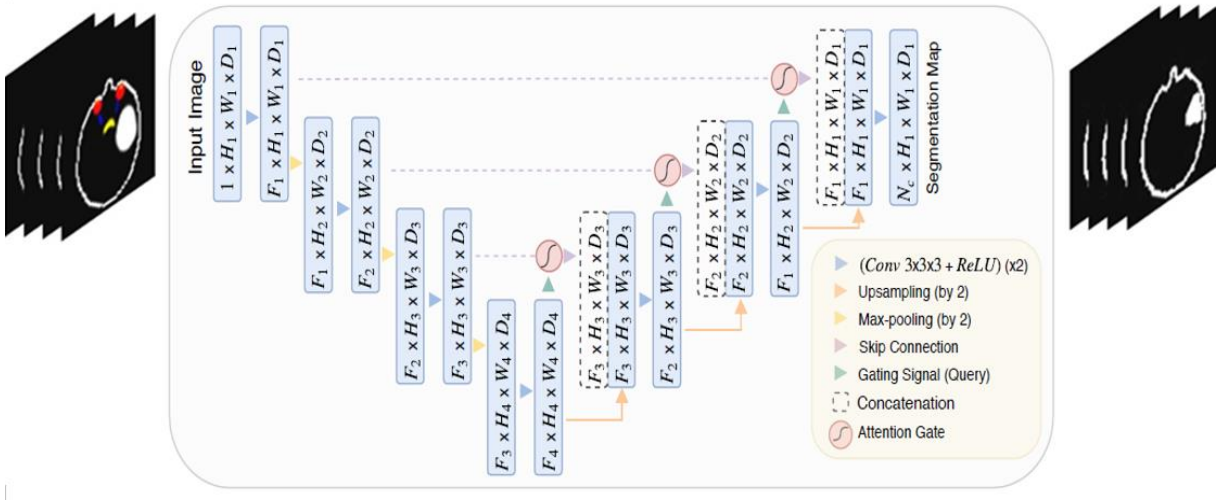

**Fig 2:** the second proposed method of network training structure

It can be seen that the input image in the encoding sector is filtered and down sampled. $N_c$ is a sample of the number of classes. Fully Convolutional Neural Networks (CNNs) perform well in traditional approaches in medical image analysis on available benchmark datasets [26, 27]. While it is faster than graph cutting and multi-atlas segmentation techniques [28].

## 2.2 Model Architecture

In this network, as mentioned before, to reduce the model's sensitivity to the foreground pixels and highlight the features transmitted through jump connections, we combined the AGs presented in 22 with model 23 in the standard U-Net architecture. In this network, we filter the neuron during the forward pass and backward pass, and we give less weight to the gradients originating from the background areas; this causes the model parameters to be in shallow layers during the backward pass. More updated based on spatial regions related to a specific task. The update rule for the convolution parameters in the $l-1$ coating can be formulated as follows:

$$\frac{\partial(\hat{x}_i^l)}{\partial(\emptyset^{l-1})} = \frac{\partial\left(\alpha_i^l f(x_i^{l-1} \emptyset^{l-1})\right)}{\partial(\emptyset^{l-1})} = \alpha_i^l \frac{f(x_i^{l-1} \emptyset^{l-1})}{\partial(\emptyset^{l-1})} + \frac{\partial(\alpha_i^l)}{\partial(\emptyset^{l-1})} x_i^l \qquad (1)$$

$$\alpha_i^l = \sigma_2(w_{att}^l(x_i^l \quad g_i \quad \theta_{att}) \qquad (2)$$

$$w_{att}^l = \gamma^T(W_x^T x_i^l + W_g^T g_i + b_g))b_\gamma \qquad (3)$$

Where Attention coefficients, $\alpha_i^l \in [0 \quad 1]$ and pixel vector $x_i^l \in \mathbb{R}^f$, a gating vector $g_i \in \mathbb{R}^f$ is used for each pixel $i$ to determine focus regions. $\sigma_2 = \frac{1}{1+\exp(-x_{i,c})}$ Correspond to sigmoid activation function, and $\theta_{att}$ is set of parameters.

To reduce the number of trainable parameters and the computational complexity of AGs, linear transformations are performed without any spatial support (1x1x1 convolution), and the input feature maps are sampled to the resolution of the gated signal similar to non-local blocks [29]. The softmax activation function normalizes the attention coefficients (2). However, sequential use of softmax results in more sparse activation at the output. For this reason, we choose a sigmoid activation function. This empirically leads to better training convergence for the AG parameters.

Corresponding linear transformations separate the feature maps and map them to a lower-dimensional space for initialization operations. As proposed in [22], low-level feature maps, first-hop connections, are not employed in the gate function because they do not show the input data in a high-dimensional space. We utilize the idea in [30] to play a role in semantically discriminative at each image scale. This helps ensure that attentional units can influence responses to a wide range of image foreground content at different scales. Therefore, we avoid reconstructing dense projections from small subsets of jump connections.

## 3. Results

Our combined AG model is modular and application-independent. It can be easily adapted for treatment dose prediction tasks for medical images. Consider the following photos to demonstrate its application. We evaluate the U-Net attention model on the treatment dose of brain cancer tumors. In particular, deformation makes determining the tumor boundary for a high percentage does reference difficult. The results of the model are shown in the figures 3, and 4, and table 2 this model is compared with the standard 3D U-Net regarding segmentation performance, model capacity, computation time, and memory requirements.

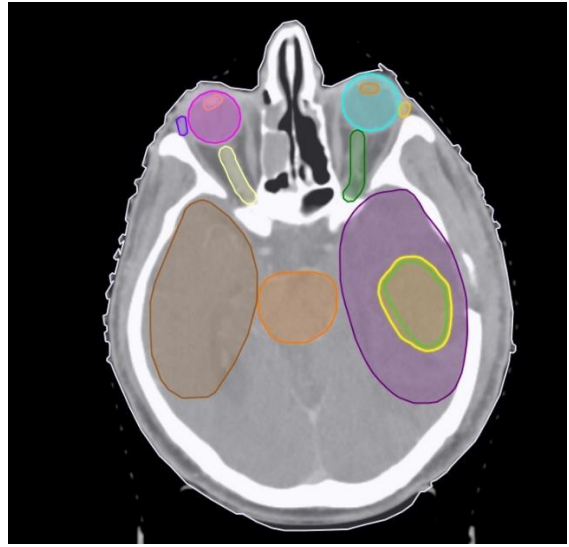

Structure

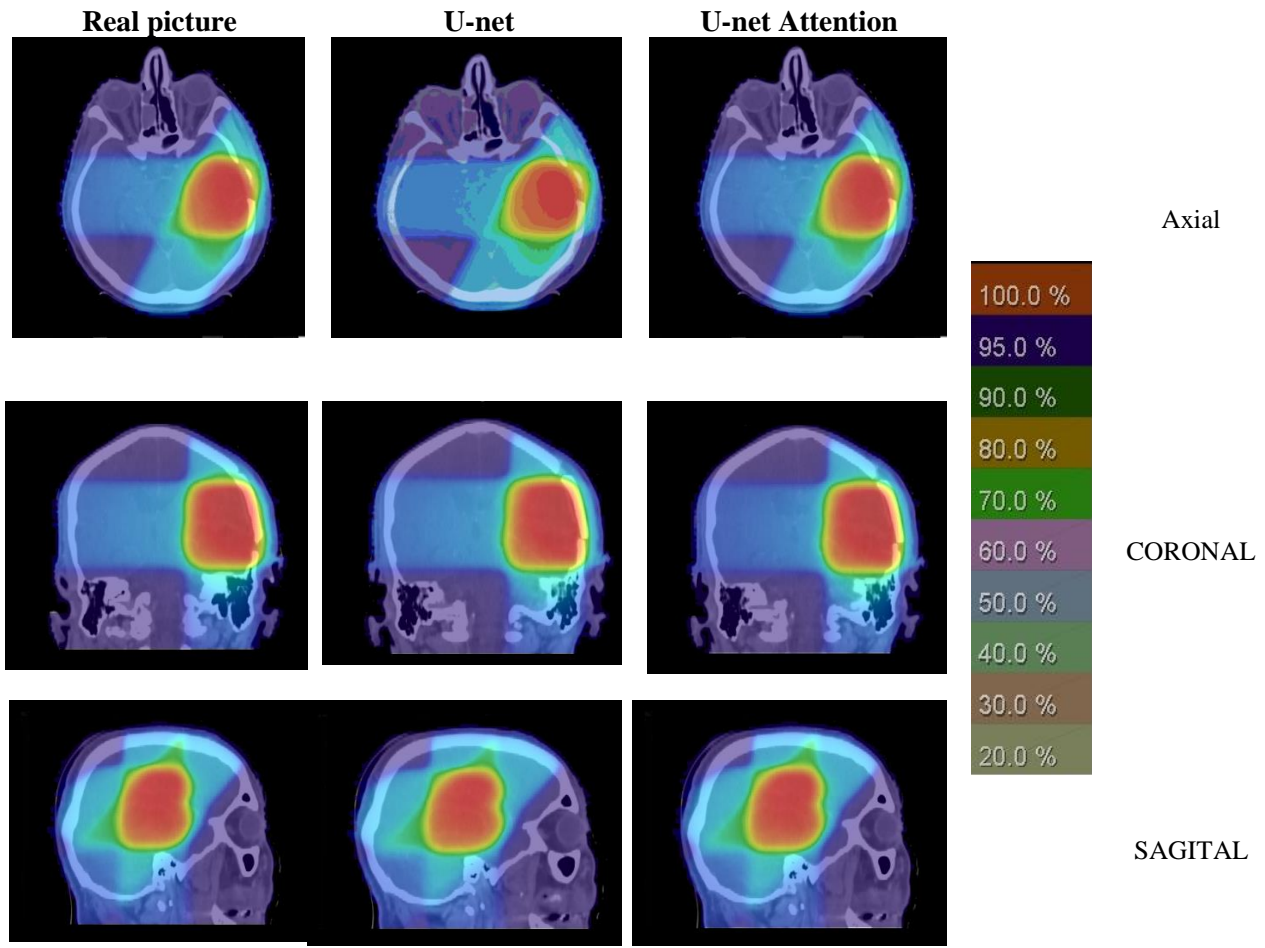

Fig 3: Visual evaluation of the predicted dose distribution in patient 93th, a) PTV±OARs (grid input image), b) desired accurate dose, c) predicted dose distribution with PTV-only method, d) predicted dose distribution with PTV-OARs method.

The images illustrate that our combined network has an acceptable performance accuracy. The proposed hybrid U-Net model has been compared with the standard U-Net in [22] for cancer tumor prediction. The training of the proposed hybrid model was run for 100 cycles, which took approximately 45 minutes for each dose percentage. We trained the network on a workstation with a 3.7 GHz Intel Core i7-8700 CPU with 32 GB RAM and a 28 GB GeForce GTX 1080 Ti graphics memory The number of patients is 9, and their images have been given to U-net Algorithms and Standard U-net algorithms for testing. Each patient has been evaluated based on different doses. The average of all doses in each patient is known. The performance of both algorithms in Only PTV and PTV-OARs methods is evaluated in Table 2. As it is known, U-Net Attention has reached an average of $97.82 \pm 0.03$ with the average of patients in the Only PTV method, which is higher than the average value of U-Net. In the PTV-OARs method, as it is known, U-Net has surpassed U-Net with an approximate value of 5.

Tables 2: comparison between U-net and Attention U-net

| methods | networks | 91* | 92* | 93* | 94* | 95* | 96* | 97* | 98* | 99* | Total Average |
|---|---|---|---|---|---|---|---|---|---|---|---|
| Only PTV | Att u-net | 97.60 ± 0.02 | 96.91 ± 0.03 | 97.95 ± 0.02 | 98.43 ± 0.02 | 95.20 ± 0.05 | 96.21 ± 0.05 | 98.13 ± 0.03 | 97.82 ± 0.05 | 96.81 ± 0.02 | **97.82 ± 0.03** |
| | U-net [22] | 95.25 ± 0.03 | 94.91 ± 0.05 | 94.39 ± 0.05 | 97.34 ± 0.03 | 92.01 ± 0.03 | 94.51 ± 0.03 | 98.30 ± 0.05 | 92.87 ± 0.02 | 95.21 ± 0.02 | 95.53 ± 0.03 |
| PTV-OARs | Att u-net | 95.25 ± 0.05 | 95.28 ± 0.03 | 94.14 ± 0.02 | 97.0 ± 0.02 | 95.74 ± 0.03 | 94.52 ± 0.02 | 98.0 ± 0.02 | 95.17 ± 0.03 | 96.0 ± 0.03 | **95.56 ± 0.05** |
| | U-net [22] | 91.45 ± 0.02 | 94.43 ± 0.05 | 86.72 ± 0.05 | 94.95 ± 0.03 | 82.0 ± 0.02 | 83.48 ± 0.02 | 95.07 ± 0.05 | 89.56 ± 0.03 | 95.02 ± 0.05 | 90.40 ± 0.05 |

Note: ∗ is the number of patients.

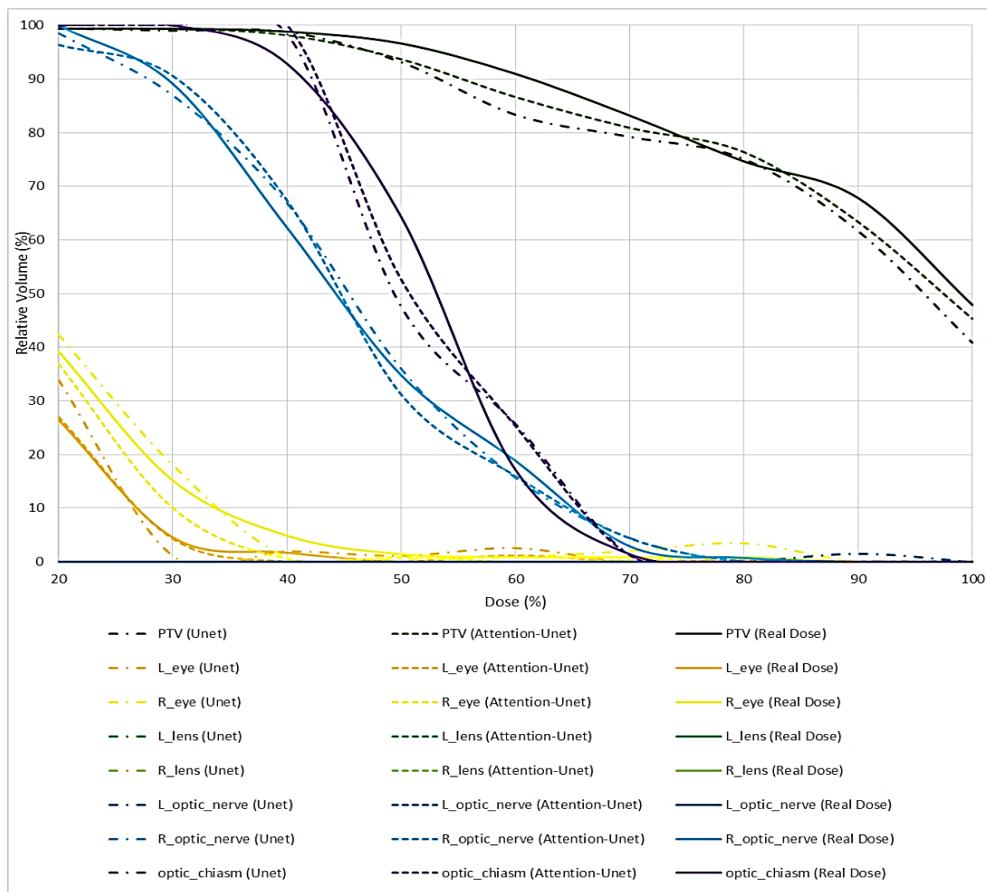

Fig 5: comparison between U-net and Att U-net with a total dose of 93th patient based on DVH, Only PTV (dot-dashed line), PTV-OARs (dashed line), Real dose (solid line).

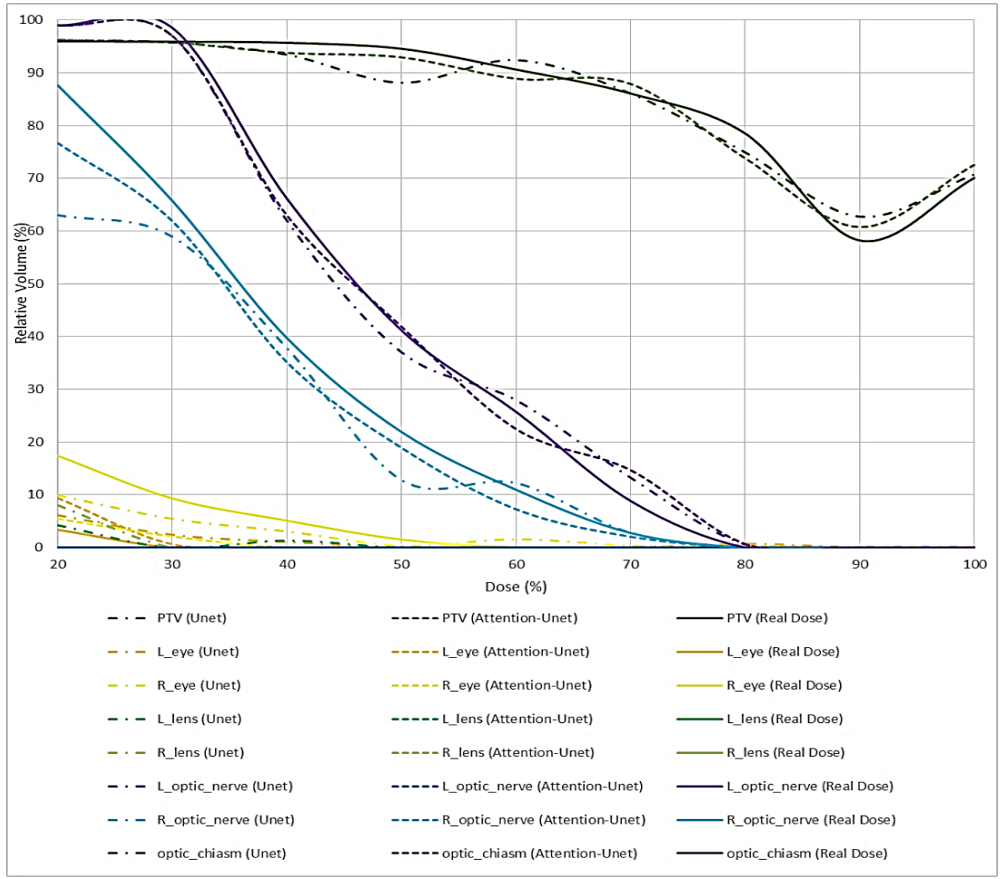

Fig 6: comparison between U-net and Att U-net with a total dose of 99th patient based on DVH, Only PTV (dot-dashed line), PTV-OARs (dashed line), Real dose (solid line).

## 4. Discussion

In this paper, we presented a combined attention gate model of 22 and the approach proposed in 34 to increase the accuracy in pixels, which is applied to predict the therapeutic dose of cancer tumors by medical images. One of the advantages of this network is eliminating the requirement to use an external object localization model. Experimental results show that combined AGs are helpful for tissue/organ identification and localization.

In this network, different learning and training plans can be used; for example, the learned weights of U-Net can be used to value the attention network. For example, this network can be used to train classification or regression used in [20] and also improve its accuracy. In this network, the sensitivity to the activity function is slightly high, and every activity function cannot train the network correctly. Researchers can discuss its adaptation in different situations in the future. On the other hand, as the disadvantages of the U-NET algorithm were mentioned in the introduction, in the results compared in the table, we saw that we could increase the accuracy to an acceptable level. In these analyses, it has been tried to use two different patients with brain tumor cancer in other areas (patient 93 in the temporal lobe (middle and front of the head) and patient 99 in the posterior cavity (back of the head).)

The presence of tumors in different areas and it's a very short distance from the OAR fig 3 has caused the treatment planning system to be focused on the design of needle distribution that has minor damage to the OARs and their surroundings, which requires an experienced person and It is highly science-aware and requires high accuracy and better analysis in artificial intelligence networks among other demanding algorithms. In Fig 4, this problem is reversed, and the treatment planning system can make a less accurate plan for distributing doses. All these discussions can be clearly described in DVH diagrams in fig 5 and 6.

Among the problems on the way, we can mention the image of a cancerous tumor in some areas of the brain because the training of the network needs different regions to have better results in accuracy for other patients under treatment. As a result, to achieve higher accuracy, separating brain regions and collecting the necessary amount of images for each category can lead to better results using the proposed methods. There is a slight difference in the results of the two proposed methods, i.e., only-PTV and PTV-OARs. We experimentally found that the prescribed dose designed by the treatment planning system is in the images containing PTV. The treatment plan is a disease-based OAR specified in the photos. These areas are not available in all cuts, so we faced a lack of data, but we were able to obtain acceptable accuracy and get a suitable dose distribution based on DVH charts.

5. Conclusion

In this article, we presented a hybrid attention gate algorithm for predicting the treatment dose of patients with brain cancer tumors. This method reduced the treatment planning time; on the other hand, we obtained better accuracy than the U-Net algorithm. Two different approaches were studied in this study, and the results were compared with the expected dose distribution. The results show the high accuracy of the proposed methods compared to the actual results based on the criteria used for evaluation. Another advantage is that the dose distribution can be predicted concisely. Therefore, using the proposed methods in the treatment planning system can improve the speed of dose distribution design. The results of the two proposed approaches, only-PTV and PTV-OARs, are that OARs can be removed from the treatment planning process, which increases the design speed.

It is suggested that researchers adapt this network to various images and the sensitivity of parameters to the randomness of values and provide an activity function to improve the network.